                                                                  \documentclass[12pt,preprint]{aastex}

\shorttitle{flux rope eruption and magnetic reconnection}
\shortauthors{Yan et al.}

\begin{document}

\title{Simultaneous observation of a flux rope eruption and magnetic reconnection during an X-class solar flare}
\author{X. L. Yan\altaffilmark{1,2,3}, L. H. Yang\altaffilmark{1,3}, Z. K. Xue\altaffilmark{1,3}, Z. X. Mei\altaffilmark{1,3},\\ D. F. Kong\altaffilmark{1,3}, J. C. Wang\altaffilmark{1,4}, Q. L. Li\altaffilmark{1,4} }

\altaffiltext{1}{Yunnan Observatories, Chinese Academy of Sciences, Kunming 650011, China. yanxl@ynao.ac.cn}
\altaffiltext{2}{Key Laboratory of Solar Activity, National Astronomical Observatories, Chinese Academy of Sciences, Beijing 100012, China.}
\altaffiltext{3}{Center for Astronomical Mega-Science, Chinese Academy of Sciences, 20A Datun Road, Chaoyang District, Beijing, 100012, P. R. China.}
\altaffiltext{4}{University of Chinese Academy of Sciences, Yuquan Road, Shijingshan Block Beijing 100049, China.}

\begin{abstract}
In this letter, we present a spectacular eruptive flare (X8.2) associated with a coronal mass ejection (CME) on 2017 September 10 at the west limb of the Sun. A flux rope eruption is followed by the inflow, the formation of a current sheet and a cusp structure, which were simultaneously observed during the occurrence of this flare. The hierarchical layers of the cusp-shaped structure are well observed in 131 \AA\ observation. The scenario that can be created from these observations is very consistent with the predictions of some eruptive models. Except for the characteristics mentioned above in the process of the flare predicted by classical eruption models, the current sheet separating into several small current sheets is also observed at the final stage of the flux rope eruption. The quantitative calculation of the velocities and accelerations of the inflow, hot cusp structure, and post-flare loops is presented. The width of the current sheet is estimated to be about 3 $\times$ $10^{3}$ km. These observations are very useful to understand the process of solar eruptions.

\end{abstract}


\section{Introduction}

The CSHKP flare model was proposed for several decades (Carmichael 1964; Sturrock 1966; Hirayama 1974; Kopp \& Pneuman 1976). In this model, magnetic reconnection of two groups of magnetic field lines with opposite directions leads to the formation of rising hot cusp-shaped coronal arcades and a current sheet during solar flares. At the same time, two bright ribbons form on either side of a prominence ( filament) or a flux rope in the chromosphere. The ribbons lie at the feet of the coronal loops and separate as the coronal loops rise. The eruption of a prominence or a flux rope is often accompanied by a CME. The prominence or the flux rope forms a bright core of a CME (Isenberg, Forbes \& Demoulin 1993; Forbes \& Acton 1996; Lin et al. 1998; Shibata \& Yokoyama 1999; Lin \& Forbes 2000; Forbes 2000; Hudson, Bougeret \& Burkepile 2006; Priest \& Forbes 2002; Lin et al. 2015).

In the past observations, a hot cusp structure in soft X-ray images was often observed during limb flares (Tsuneta et al.1992; Tsuneta 1996). Moreover, three hard X-ray sources were found at the maximum of solar flares. One of them is situated at the top of the cusp structure and the other two ones were located at the footpoints of the cusp structure (Masuda 1994; Sui \& Holman 2003; Sui, Holman, \& Dennis 2004). Direct observation of a long current sheet after a CME was reported by Lin et al (2005). Besides, a clear reconnection inflow and X-point were found in a limb flare on 1999 March 18 (Yokoyama, Yamamot \& Fukao 2001). With the development of the observational technology, more and more features of magnetic reconnection were reported (Savage et al. 2010; Liu et al. 2010; Zhang et al. 2012; Sun et al. 2012;  Savage et al. 2012a, 2012b; Su et al. 2013; Zhang et al. 2013; Deng et al. 2013; Dud{\'{\i}}k et al. 2014; Tian et al. 2014; Yang, Zhang \& Xiang 2015; Sun et al. 2015; Xue et al. 2016; Li et al. 2016; Zhu et al. 2016; Li et al. 2017; Shen et al. 2017). Furthermore, many simulations also show evidence that magnetic reconnection plays an important role in solar eruptions (Antiochos et al. 1999, Chen \& Shibata 2000, Amari et al. 2003, Aulanier et al. 2010, Janvier et al. 2013; Ni et al. 2016, 2017; Mei et al. 2017; Jiang et al. 2016, 2017). 

In this letter, we present a relatively rare observation of a complete eruptive process of an X8.2 flare associated with a CME in active region NOAA 12673 on 2017 September 10. The details of the observations and methods are presented in section 2. The results are shown in section 3. Conclusion and discussion are given in section 4.

\section{Observations and Methods}

The Atmospheric Imaging Assembly (AIA; Lemen et al. 2012) on board the Solar Dynamics Observatory (SDO) can provide 10 wavelengths of full-disk solar atmospheric images. The spatial and temporal resolutions of the AIA data are 1.$^\prime$$^\prime$2 and 12 s, respectively. The 211 \AA, 171 \AA, 131 \AA\ images are employed to show the  eruption process of the X8.2 flares. All the SDO data are calibrated to Level 1.5 by using the standard procedure in Solar Software (SSW) package.

Data form the Reuven Ramaty High Energy Solar Spectroscopic Imager (RHESSI) (Lin et al. 2002) were also used to study the X-ray sources during the X8.2 flare. The energy ranges of 6 - 12 keV and 12 - 25 keV were selected to identify the X-ray sources during the flare. The accumulation time is 30 s for constructing X-ray images by using the Clean imaging algorithm.

The differential emission measure (DEM) was calculated by using the almost simultaneous observations of six AIA EUV lines (e.g., 131 \AA, 94 \AA, 335 \AA, 211 \AA, 193 \AA, and 171 \AA\ formed at coronal temperatures) observed by SDO/AIA. The observed intensity $I_{i}$ for each wavelength can be determined by 
\begin{equation}
I_{i} = \int R_{i}(T) \times DEM(T)dT,
\end{equation}
where $I_{i}$ is the observed intensity of the waveband $i$ and  $R_{i}(T)$ represents the temperature response function of wavelength $i$. $DEM(T)$ is the DEM of coronal plasma, which is computed using the routine \texttt{xrt\_dem\_iterative2.pro} in the SSW package. This code was first written for processing Hinode X-ray Telescope data (Golub et al. 2004; Weber et al. 2004), and then modified for handling SDO/AIA data (Schmelz et al. 2011; Cheng et al. 2012). In this letter, log $T$ is set in the range of 5.7--7.3, where the DEM is generally well constrained (Aschwanden \& Boerner 2011, Hannah \& Kontar 2012). 

To obtain the emission measure (EM) in the temperature ranges $[T_\mathrm{min},\,T_\mathrm{max}]$, we evaluate EM using the following equation:
\begin{equation}
EM=\int _{T_{min}}^{{T}_{max}}DEM(T)dT. 
\end{equation}

The DEM-weighted (mean) temperature is calculated as follows:

\begin{equation}
T_{mean}=\frac{\int _{T_{min}}^{{T}_{max}}DEM(T)TdT}{\int _{T_{min}}^{{T}_{max}}DEM(T)dT}. 
\end{equation}

\section{Result}
Active-region NOAA 12673 rotated at the west solar limb on 2017 September 10. It produced an X8.2 flare. The flare started at 15:35UT, peaked at 16:06 UT, and ended at 16:31 UT. This flare is also associated with a CME.

Figure 1 shows the process of the X8.2 flare in 171 \AA, 211 \AA, and 131 \AA\ observation from SDO. At the beginning of the flare, there is a flux rope with a filament in it. The filament can be seen in SDO 304 \AA\ observation from 15:32 UT to 15:53 UT. The filament is marked by the blue arrows in Figs. 1(a) and 1(e). At about 15:30 UT, the flux rope began to rise slowly. The green arrows in Figs. 1(b), 1(f), and 1(j) indicate the location of the flux rope. The flux rope has a dark cavity and bright outer ring. The cavity is much clearer in 211 \AA\ image (sensitive to a plasma temperature of $\sim$ 2 MK) than in the other passbands. The bright outer ring of the flux rope can be seen in 131 \AA\ (see Fig. 1(j)) and 94 \AA\ (sensitive to a plasma temperature of $\sim$ 7 MK, not shown in the paper) observations. Therefore, the flux rope is corresponding to a very hot channel (Zhang, Cheng \& Ding 2012; Li \& Zhang 2013; Cheng et al. 2014). After the eruption of the flux rope, there is a long current sheet behind it. The red arrows in Figs. 1(c), 1(g), and 1(k) indicate the current sheet. Moreover, a small cusp structure formed at the lower solar atmosphere. At the final stage of the flare, the post-flare loops can be seen clearly in 171 \AA\ and 211 \AA\ observations (see the pink arrows in Figs. 1(d) and 1(h)). {\bf A well-defined cusp structure with hierarchical structure appeared in 131 \AA\ observation and its inner layer is brighter than the outer one (see the pink arrow in Fig. 1(l)).}

Composite images are presented in Fig. 2. The red, green, and blue colors indicate the 131 \AA, 211 \AA, and 171 \AA\ observations (see Figs. 2(a)-2(d)). The yellow arrows indicate the filament, flux rope, current sheet, and cusp structure, respectively. The process of the X8.2 flare can be seen in animation 1. The flux rope experienced a slow rise from about 15:30 UT to 15:46 UT and then began to accelerate at about 15:46 UT. During its rising, the flux rope itself became larger and larger due to its expansion. To evaluate the velocities and accelerations of the flux rope, inflow, cusp structure, and post-flare loops, two time slices are made along slit A-B and C-D. The white lines in Fig. 2(e) indicate the positions of the two slits. The time slice shown in Fig. 2(f) is made along the slit A-B direction. The blue dotted lines 1 and 2 in Fig. 2(f) are used to evaluate the average velocity of the inflow. From the time slice, it is very clear that the inflow appeared after the flux rope at about 15:52 UT. Note that we only used 171 \AA\ images to make a time slice to obtain the projection inflow velocity (The time slice made by using the composite images is not so clear). The average velocities of the inflows from two sides to the reconnection region are about 111.6 $\pm$ 4.5 km $s^{-1}$ and 104.5 $\pm$ 3.9 km $s^{-1}$, which are calculated along lines 1 and 2 in Fig. 2(f) by using linear fitting. To determine the inflow velocities, we track the loops in 171 \AA\ observation and find their corresponding positions in the time slice. We outline the path of the inflow and repeated 10 times to get the uncertainty error bars. The inflow velocities evaluated in the letter are much larger than that of the inflows observed in some previous literatures (e.g., several km $s^{-1}$, Yokoyama et al. 2001; Sun et al. 2015; Xue et al. 2016) and a little smaller than the inflows from 150 to 690 km $s^{-1}$ reported by Savage et al. (2012a). Comparing the event in this letter and that studied by Savage et al. (2012a) with other events, we found that the inflow in the events with an erupting flux rope is distinctly larger than that without an erupting flux rope. 

The time slice shown in Fig. 2(g) is made along the slit C-D direction (see the white solid line in Fig. 2(e)). The red dashed line in Fig. 2(g) outlines the slow rise stage of the flux rope from about 15:30 UT to 15:46 UT. The average velocity is about 5.3 $\pm$ 0.4 km $s^{-1}$. The white dashed curve lines 3, 4, and 5 outline the acceleration stage of the top of the flux rope, the bottom of flux rope, and the the cusp-shaped structure. The velocities and accelerations of them are derived along these lines. We used quadratic fitting for lines 3, 4, and 5.  The average velocity of the flux rope is evaluated along lines 3 and 4. Line 3 indicates the upper of the flux rope and line 4 indicates the bottom of the flux rope. The average rising velocities and accelerations of the upper (bottom) part of the flux rope are 224.7 $\pm$ 2.8 km $s^{-1}$ (191.9 $\pm$ 4.0 km $s^{-1}$), 1.4 $\pm$ 0.04 km $s^{-2}$ (1.3 $\pm$ 0.09 km $s^{-2}$), respectively. The average velocity and acceleration of the cusp structure are 65.9 $\pm$ 2.7 km $s^{-1}$ and 0.3 $\pm$ 0.02 km $s^{-2}$, which are calculated along line 5 in Fig. 2(g). The dashed black line 6 outlines the time distance diagram of the post-flare loops. We use  the linear fitting for line 6. The average velocity of the post-flare loops is 35.4 $\pm$ 3.0 km $s^{-1}$ , which is calculated along line 6 in Fig. 2(g). Similar to the inflow velocities, the uncertainty error bars of the velocities of the flux rope and the cusp structure are estimated through 10 times of repeated measurement. Note that there are certainly errors or biases beyond those from manually tracking the inflow/flux rope working with EUV images to get the velocities. Moreover, the images suffer from projection effect and propagation away from the slit. From the time slice, there is an obvious acceleration of the rising velocity of the flux rope before the occurrence of the magnetic reconnection. At about 15:53 UT, the cusp-shaped structure began to rise and accelerate. The fast rising of the flux rope was ahead of the magnetic reconnection. The rising of the cusp-shaped structure and the occurrence of magnetic reconnection occurred almost simultaneously. The observations fit very well with the reconnection model of CME eruption.

Figure 3 shows the eruption of the flux rope and the formation of the current sheets and cusp structure in 131 \AA\ observations. The red dashed lines outline the flux rope (see Figs. 3a-3c). The two legs of the flux rope rooted in the active region with a dip at its center part. The dip rose slowly and developed into a dark cavity shown in Figs. 1(b), 1(f), and 1(j). The cavity is corresponding to the center part of a cross-section of the flux rope. After the eruption of the flux rope, a long current sheet appeared behind the flux rope (see the yellow arrow in Fig. 3d). {\bf The inner layers of the cusp structure are much brighter than outer layers (see Fig. 3(f))}. At the final stage, several small current sheets appeared near the original one (see the yellow arrows in Figs. 3e-3f). Note that the small current sheets can be only seen in 131 \AA\ observations, which is sensitive to a plasma temperature of $\sim$ 0.4 and $\sim$ 11 MK. Up to now, there is no a conclusive answer on the width of the current sheet. We measure the width of the current sheet along the white line that crosses the current sheet (see the white line in Fig. 3(d)). The intensity profile (red points) is shown in Fig. 3(d). The average intensity of ten points perpendicular to the white line is calculated to make the profile plot (5 point of each side of the white line). A Gaussian fitting is used to fit the intensity profile (see the green line). The full width at half maximum value of the Gauss profile is taken as the width of the current sheet. The width of the current sheet is estimated to be about 3.3 $\times$ $10^{3}$ km, which approaches the width of the current sheet obtained by Liu et al. (2010) ( (5-10) $\times$ $10^{3}$ km) and Savage et al.(2010) ((4-5) $\times$ $10^{3}$ km). The width of the current sheet agrees with the prediction of reconnection theory (Priest \& Forbes 2000). But it is much narrower than that estimated by Ko et al. (2003), Webb et al. (2003), Lin et al. (2007), and Ciaravella \& Raymond (2008).The detailed process of the X8.2 flare in 131 \AA\ observation can be seen in animation 2.

The X-ray sources of 6-12 keV (red contours) and 12-25 keV (blue contours) were overlaid on the 131 \AA\ images in Figs. 4(a) and 4(b). The contour levels are the 10\%, 40\% and 80\% of the maximum value of 6-12 keV and 12-25 keV. The contours of the sources cover the whole cusp structure. The maximum points are located at the top of the cusp structure. At the original stage of the flux rope eruption, the positions of 6-12 keV and 12-25 keV X-ray sources are almost overlapping together. The cusp structure became larger following the new formed post-flare loops. The center of the ellipses where 6-12 keV and 12-25 keV X-ray sources are about 22 and 27 arcseconds higher than before, respectively. These X-ray sources are located at the top of the cusp-shaped structure. The hard X-ray source is closer to the top of the cusp-shaped structure than the soft X-ray source. Using the method described in the section of Observations and methods, we calculate the EM at the different temperature bins. Figs. 4(c), 4(d), and 4(e) show the EM maps at the different temperature bins at 13:53:30 UT (Cheng et al. 2012). It is clear that the center of the flux rope has lower EM and lower temperature. The edge of the flux rope has relatively higher EM and higher temperature. The bottom of the flux rope exhibits higher EM, where there may be have largher plasma density than the top of the flux rope. The temperature map is shown in Fig. 4(f). It is very clear that the edge of the flux rope has relatively higher temperature (about 10 MK) (see Fig. 4(f)). It is confirmed that a flux rope is a hot channel.

According to the configuration of the eruptive structure, a cartoon is drawn on the composite image (see Fig. 5). The dark cavity surrounding with a bright ring represents a flux rope. The overlying magnetic field lines enclosed it. After the rising of the flux rope, a cusp structure forms at the bottom of the solar atmosphere. Moreover, a long current sheet forms between them. This picture is very consistent with the models proposed by Lin \& Forbes (2000) and Shibata (1999).

\section{Conclusion and discussion}
The whole process of an X8.2 flare was covered by the observation of SDO on 2017 September 10. At the beginning of the flare, a flux rope with a filament rose rapidly. Behind the flux rope, a very long current sheet formed due to the magnetic reconnection of overlying magnetic field lines of the flux rope. At the same time, a cusp structure appeared at the bottom of solar atmosphere. At the final stage, several small current sheets appeared near the original one at 131 \AA\ observation. This event is clear to exhibit a complete eruptive process of a solar eruption. Moreover, this event is associated with an X8.2 flare and a CME. Therefore, it is a very typical event for studying solar eruptions. All the observations are very consistent with the predictions of the previous solar eruption models (e.g., CSHKP model, Lin \& Forbes 2000, and Shibata et al. 1999). Through the study of this event, it is very helpful to understand the nature of solar eruptions. Furthermore, the detailed velocities and accelerations of the rising of the flux rope,  cusp structure, post-flare loops are very useful to set up a complete solar eruption model. 

Using SOHO/EIT 195 \AA\ observation, Yokoyama et al. (2001) found an apparent magnetic reconnection inflow of 5 km $s^{-1}$ in a limb flare. The similar result was reported by Sun et al. (2015) and Zhu et al. (2016). The inflow obtained by Su et al. (2013) and Wang et al. (2017) is about tens of km $s^{-1}$.  A relatively large inflow from 150 to 690 km $s^{-1}$ was found in a solar flare by Savage et al. (2012). The events studied by Savage et al. (2012) was associated with an erupting flux rope. The inflow found in this letter is about 100 km $s^{-1}$. Similar to the event of Savage et al. (2012), the inflow is followed by the eruption of a flux rope. Consequently, the fast inflow may be associated with the acceleration eruptions of flux ropes. {\bf The formation of the cusp-shaped structure has an acceleration phase, which started at the same time as the appearance of the signatures of magnetic reconnection.} Therefore, the acceleration rising of the cusp-shaped structure is found to be very closely related to the occurrence of magnetic reconnection signatures. The post-flare loops also have a stable rising velocity. With the rising of the flux rope, more and more magnetic field lines moved into the reconnection region and reconnected together. The width of the current sheet estimated in this letter is about 3 $\times$ $10^{3}$ km, which is similar to the results obtained by Liu et al. (2010), Savage et al. (2010), and Zhu et al. (2016). But it is at about 10 times thinner than the results obtained by Ko et al. (2003), Webb et al. (2003), Lin et al. (2007), and Ciaravella \& Raymond (2008) by using UVCS and SOHO/LASCO data. Furthermore, the flux rope is confirmed to be corresponding to a hot channel by using DEM method as the results obtained by Zhang et al. (2012) and Cheng et al. (2014). The temperature of the hot channel is about 10 MK. The eruption of the flux rope experienced a slow rise and a fast rise phase as well as that of CMEs (Zhang et al. 2001). At the late stage of the flux eruption, several small current sheets formed. It seems the main current sheet was separated into the small current sheets that is not predicted by the current eruption models.

\acknowledgments
The authors thank the referee for her/his constructive suggestions and comments that helped to improve this paper. We would like to thank the SDO/AIA and RHESSI teams for the high-cadence data support. This work is sponsored by the National Science Foundation of China (NSFC) under the grant numbers (11373066, 11603071, 11503080, 11633008, 11763004, 11533008, 11303088), by the Key Laboratory of Solar Activity of CAS under numbers KLSA201603, KLSA201508, by the Youth Innovation Promotion Association CAS (No.2011056), and by the grant associated with project of the Group for Innovation of Yunnan Province.

\begin{figure}
\epsscale{.7}
\plotone{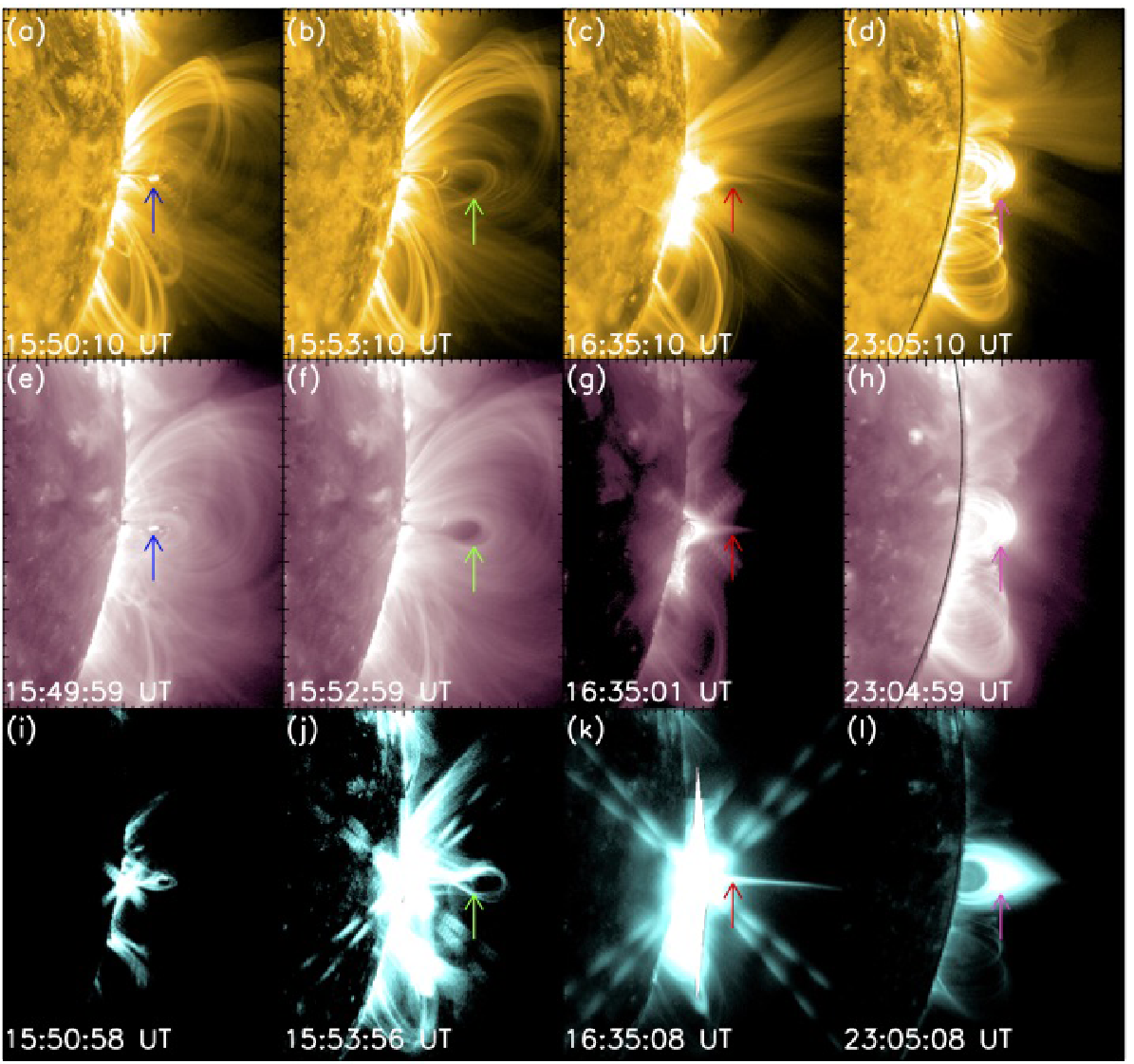}
\caption{Process of the X8.2 flare in 171 \AA, 211 \AA, and 131 \AA\ observations from SDO. The filament is marked by the blue arrows in Figs. 1(a) and 1(e). The green arrows in Figs. 1(b), 1(f), and 1(j) indicate the location of the flux rope. The red arrows in Figs. 1(c), 1(g), and 1(k) indicate the current sheet. The pink arrows in Figs. 1(d) and 1(h) indicate the post-flare loops, while the pink arrow in Fig. 1(l) indicate the cusp structure at 131 \AA\ observations.\label{fig1}}
\end{figure}

\begin{figure}
\epsscale{.8}
\plotone{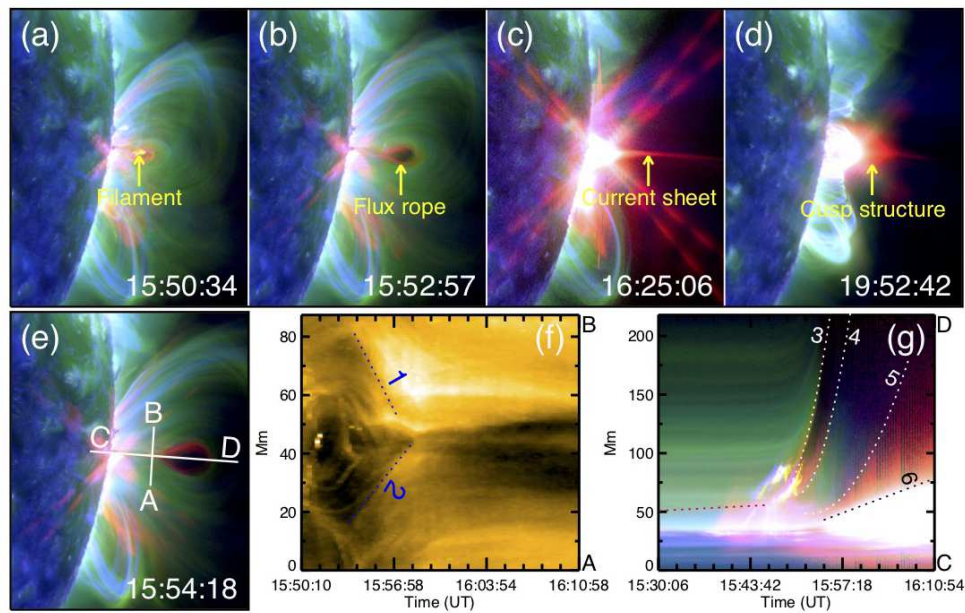}
\caption{Composite images to show the process of the X8.2 flare. The red, green, and blue colors indicate the 131 \AA, 211 \AA, and 171 \AA\ observations. The yellow arrows in Fig. 2(a)-(d) indicate the filament, flux rope, current sheet, and cusp structure, respectively. The white lines in Fig. 2(e) denote the slits that are used to make the time slices of Fig. 2(f) and 2(g). The average inflow velocities are derived along the dashed lines 1 and 2 in Fig. 2(f). The average velocity during the slow rise of the flux rope is derived along the red dashed line in Fig. 2(g). The average velocities and acceleration of the flux rope, and cusp structure are  derived along the white dashed curve lines 3, 4, and 5 and the average velocity of the post-flare loops is calculated along the dashed line 6 in Fig. 2(g). An animation (animation 1) with composite images is available. \label{fig1}}
\end{figure}

\begin{figure}
\centering
\includegraphics[angle=0,scale=0.8]{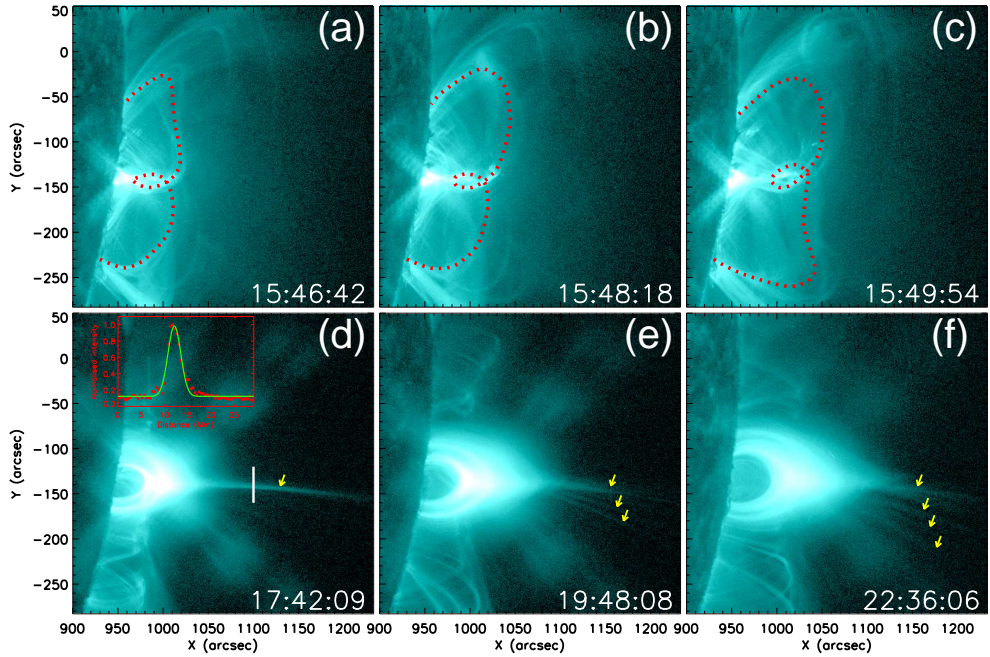}
\caption{Eruptive process of the X8.2 flare at 131\AA\  observation. The red dashed lines outline the shape of the flux rope. The yellow arrows indicate the current sheets. The intensity profile in Fig. 3(d) is calculated along the white line that crosses the current sheet. The full width at half maximum of the Gauss profile is take as the width of the current sheet. An animation (animation 2) with 131\AA\ images is available.}
\end{figure}

\begin{figure}
\epsscale{.850}
\plotone{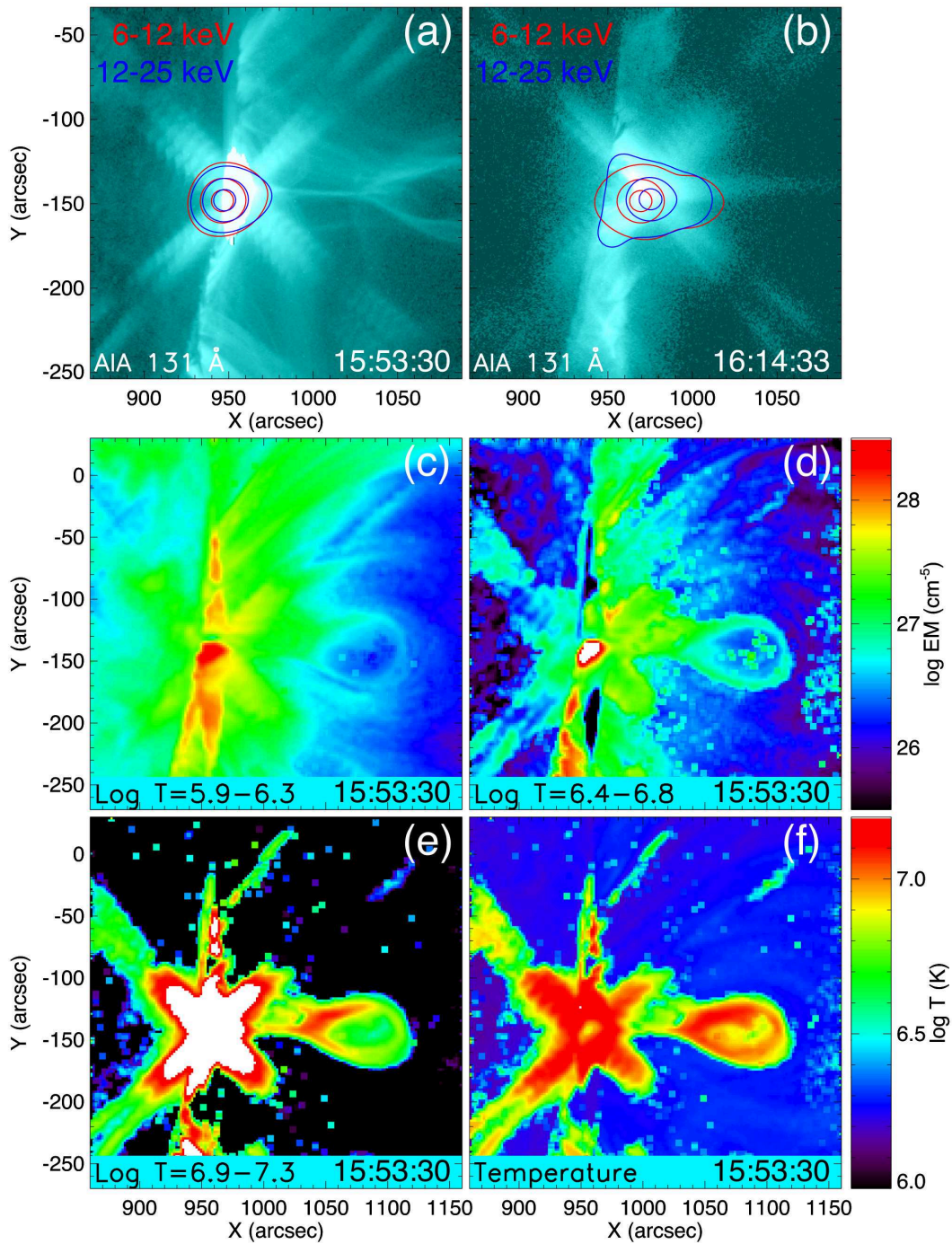}
\caption{RHESSI X-ray sources of 6 - 12 keV (red lines) and 12 - 25 keV (blue lines) overlaid on the 131 \AA\ images, DEM map at different temperature band, and temperature map derived from the DEM method. The contour levels are the 10\%, 40\% and 80\% of the maximum value of 6-12 keV and 12-25 keV.\label{fig1}}
\end{figure}

\begin{figure}
\epsscale{.6}
\plotone{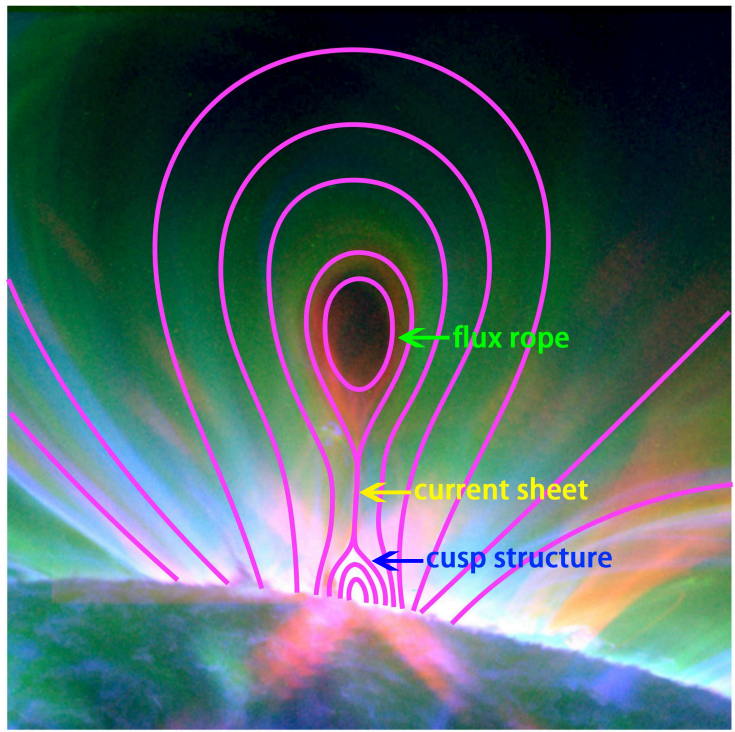}
\caption{A cartoon drawn on the composite image to show the eruptive structure during the X8.2 flare. The pink lines outline the flux rope, current sheet, cusp structure, and overlying magnetic field lines. \label{fig1}}
\end{figure}


\end{document}